\documentclass[aps,prl,twocolumn,groupedaddress,showpacs,floatfix]{revtex4}
\usepackage{graphicx}
\usepackage{psfig}
\usepackage{epsf}
\usepackage{epsfig}

\begin{document}

\title{Coulomb crystallization in expanding laser-cooled neutral plasmas}

\author{T.\ Pohl}
\author{T.\ Pattard}
\author{J.M.\ Rost}

\affiliation{MPI for the Physics of Complex Systems, N{\"o}thnitzer
Str.\ 38, D-01187 Dresden, Germany}

\date{\today}

\begin{abstract}
We present  long-time simulations of expanding ultracold neutral
plasmas, including a full treatment of the strongly coupled ion dynamics.
Thereby, the relaxation dynamics of the expanding
laser-cooled plasma is studied, taking into account elastic as well as
inelastic collisions.
It is demonstrated that,
depending on the initial conditions, the ionic component of
the plasma may exhibit short-range order or even a superimposed
long-range order resulting in concentric ion shells.
In contrast to ionic plasmas confined in traps, the shell structures are built
up from the center of the plasma cloud rather than from the periphery.
\end{abstract}


\maketitle
It is well known that,
depending on the Coulomb coupling parameter (CCP) $\Gamma=e^2/a k_B T$,
a plasma may show long-range order,
short-range order or no order at all. In general, ordering
effects can be expected
in the so-called strongly coupled regime ($\Gamma\gg 1$) 
where the interparticle Coulomb interaction $e^2/a$ dominates the
thermal energy $k_B T$ of the plasma particles.
This parameter regime has been studied extensively
in nonneutral plasmas of laser-cooled ions confined in ion traps
\cite{Rah86,Dub88,Gil88,Dub99,Tot02}. On the other hand, much less is known
about the dynamics of finite, strongly coupled neutral plasmas without
confinement.
Due to their expansion, these plasmas are in a non-equilibrium state at all
times, and it is not clear whether dramatic ordering effects such as Coulomb
crystallization known from static trapped ionic plasmas can be observed
in such a system.

Experimentally, cold neutral plasmas could be realized only recently by
photoionizing a cloud of ultracold ($T\ll 1$K) atoms \cite{Kil99,Kul00,Kil01}.
Yet, under the present experimental conditions, the regime
of a strongly correlated plasma cannot be reached \cite{Mur01,Kuz02,Rob02}. 
Since the plasma is created in a completely uncorrelated
state, the subsequent conversion of potential into kinetic energy rapidly heats both the
electron and ion subsystem, suppressing the development of substantial
correlations \cite{Mur01,Kuz02}. Additionally,
inelastic collisions with Rydberg atoms, previously formed by
three-body recombination (TBR), and the TBR itself
\cite{Rob02,Rob03} heat the electron gas. Therefore, $\Gamma_i$ decreases
to unity while $\Gamma_e$
becomes even smaller, preventing strong correlation effects.
However, further Doppler cooling of the ions
during the plasma expansion  has been suggested as a possible route to
strongly coupled ultracold plasmas \cite{Kuz02b,Kil03}.

In the following we provide the first theoretical description for the expansion
of such a laser-cooled neutral plasma. As we will demonstrate, strong-coupling 
phenomena can indeed occur in such a system, leading to the formation of
surprisingly differentiated patterns in the ionic plasma
given the appropriate initial conditions. From a nonlinear dynamics point of
view one might characterize these phenomena as self organization  of a system in
a non-equilibrium state.
 
In a first step we describe the collisionless plasma
dynamics by a set of coupled Vlasov equations for the electrons
and ions \cite{Dor98}, neglecting any correlation effects. Cooling of the ions
is modelled by introducing a Fokker-Planck term into the ion kinetic equation
\cite{Met99}
\begin{equation} \label{fp}
\left(\frac{\partial f_i}{\partial t}\right)_c=\beta\left[\nabla_{\mathbf{v}}
\left({\mathbf{v}}f_i\right)+\frac{k_BT_c}{m_i}\Delta_{\mathbf{v}}f_i\right]
\;\; .
\end{equation}
Here, $f_i$ is the one-particle distribution function of the ions, $m_i$ is
the ion mass and the damping rate $\beta$ and the Doppler temperature $T_c$
are determined by the details of the laser cooling process.
Assuming quasineutrality together with an adiabatic treatment of the electrons,
one can show that the resulting ionic kinetic
equation permits a Gaussian selfsimilar solution
$f_i\left({\bf{r}},{\bf{v}}\right)\propto\exp(-{r^2}/{2\sigma^2}-{m_i\left({\bf{v}}-\gamma{\bf{r}}\right)^2}/{2k_B
T_i})$, also found in the free plasma expansion problem
\cite{Dor98,Rob02,Poh03}.
The rms-radius $\sigma$ of the plasma cloud, the hydrodynamic velocity parameter
$\gamma$ and the temperatures $T_i$ and $T_e$ evolve according to
\begin{eqnarray} \label{quneutr}
\dot{\sigma}&=&\gamma\sigma, \nonumber \\
\dot{\gamma}&=&\frac{k_BT_e+k_BT_i}{m_i\sigma^2}-\gamma\left(\gamma+\beta\right),
\nonumber \\
\dot{T}_e&=&-2\gamma T_e\:,\quad\dot{T}_i=-2\gamma T_i-2\beta\left(T_i
-T_c\right).
\end{eqnarray} 
As seen from Eq.~(\ref{quneutr}), the action of the cooling laser is
twofold. The ion temperature is driven towards its equilibrium value $T_c$,
while $\gamma$ is linearly damped out on a timescale of $\beta^{-1}$.
Eq.~(\ref{quneutr}) still has one integral of motion, $\sigma^2T_e=\mathrm{const.}$,
reflecting the adiabatic electron cooling during the plasma expansion.
If $T_i \ll T_e$, an asymptotic description of the plasma dynamics can be
obtained by neglecting terms of order $\gamma$ compared to the damping rate
$\beta$ in Eq.~(\ref{quneutr}), which yields in the long-time limit
\begin{equation} \label{asympt}
\sigma^2\approx\sigma^2(0)\sqrt{1+2\frac{k_BT_e(0)}{\beta
m_i\sigma^2(0)}t} \quad \stackrel{t\rightarrow \infty}{ \propto} \sqrt{t} \;\;.
\end{equation}
This is in marked contrast to the free expansion, where
$\sigma^2\approx\sigma^2(0)+\frac{k_BT_e(0)}{m_i}t^2 \stackrel{t\rightarrow \infty}{ \propto} t^2$. Therefore,
continued Doppler cooling of the ions
not only reduces their temperature but also drastically retards the decrease
in plasma density, supporting development of strong correlations.

In a second step, we have performed a more elaborate numerical simulation
based on a hybrid method treating the two plasma components
on different levels of sophistication.
Since the electrons are not strongly coupled
and their relaxation time, determined by the electron plasma frequency,
is small compared to both the plasma expansion time and the inverse of the
ionic plasma frequency $\omega_p = \sqrt{4\pi e^2\rho/m_i}$, an equilibrium
fluid model provides an adequate description of the electron dynamics \cite{Rob02}.
We account for the initial electron evaporation by determining the
fraction of trapped electrons from the results of Ref.~\cite{Kil99}.
The
ions  move as classical particles under the action of the electronic
mean field and the direct ion-ion interaction calculated with a particle
tree-procedure developed in \cite{Bar86}. In analogy to
Eq.~(\ref{fp}), cooling is described by adding a Langevin force to the ion
equations of motion. The electron temperature is determined by
energy conservation for the total system consisting of the plasma and the
radiation field. This hybrid treatment allows
us to study effects of strong ion correlation over long times, since atomic timescales 
need not be resolved as in a full molecular dynamics (MD) simulation
\cite{Kuz02,Maz02}.
\begin{figure}[tb]
\centerline{\psfig{figure=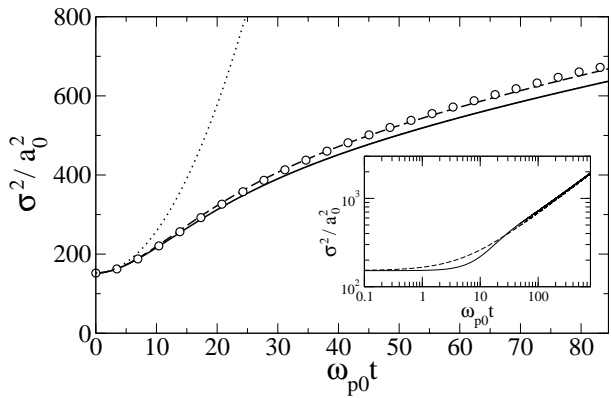,width=8cm}}
\caption{\label{fig1}
The size of a laser-cooled plasma as compared to that of a freely expanding
plasma of the same initial-state parameters (dotted). The solution of
Eq.\ (\ref{quneutr}) (dashed) matches the PIC-treatment (circles), showing that
the difference to the hybrid-MD simulation (solid) comes from the ionic
correlations. The inset shows a comparison of the numerical solution with
the analytical approximation Eq.\ (\ref{asympt}).}
\end{figure}

In Fig.~\ref{fig1}, we compare the time dependence of the plasma rms-radius obtained from Eq.~(\ref{quneutr}) with the hybrid-MD
simulation and the analytical approximation Eq.~(\ref{asympt}).
There is good overall agreement between the two numerical approaches.
Moreover, they both nicely reproduce the asymptotic $\sqrt{t}$-behavior of
Eq.~(\ref{asympt}). On the other hand, the width calculated from the
MD simulation is significantly shifted to lower values.
A comparison with a
particle-in-cell (PIC) treatment of the ions, also shown in Fig.~\ref{fig1},
clearly reveals that the slower plasma
expansion is due to the negative correlation pressure \cite{Kuz02,Poh03},
which partly compensates the thermal electron pressure. Note that here the
influence of ion correlations is completely different from the case of free
plasma expansion, where the initial correlation heating was found to dominate
the negative correlation pressure and hence accelerates the plasma expansion
\cite{Poh03}.

Up to this point, we have taken into account the electron-ion interaction on
the basis of a mean field description only. However, it has been found that
electron-ion collisions leading to the formation of Rydberg atoms through
TBR
may significantly alter
the expansion dynamics at these low electron temperatures \cite{Kil99,Rob02}.
In order to include these processes in our description, we use a
Monte-Carlo treatment \cite{Vah95,Rob03} to account for TBR and
inelastic electron-Rydberg atom collisions.
In addition to these processes 
the influence of the cooling laser on the
Rydberg atoms should be addressed. By the very nature of the cooling process,
a significant fraction of the ions is found in an excited state at all times.
Thus, TBR may produce doubly-excited, and hence autoionizing,
Rydberg atoms with a considerably large rate. (This
process is in close analogy to the production of autoionizing Rydberg states
by ``isolated-core excitation'' \cite{Gal94}.)
For low enough principal quantum numbers, the autoionization rate of these
states becomes comparable to and even exceeds the radiative decay rate of the
excited core.
For the case of strontium, the l-averaged autoionization rate
becomes important at $n\approx 50$. For electron temperatures of the present
type of experiments, Rydberg atoms typically recombine
into states with much higher $n$ so that Auger ionization
does not play a  role initially. However, in the course of the gas evolution the
Rydberg electron moves down the energy ladder by subsequent inelastic
electron-atom collisions.
Since the energy
shift of the core transition used for laser cooling is of the same order of
magnitude as the autoionization rate \footnote{For the
specific case of strontium, the energy shift and the autoionization rate were
found to be almost identical.}, the Rydberg atoms formed by TBR are still
resonant with the cooling laser. Hence, even if the timescale of collisional
deexcitation is longer than the lifetime of the core-excited state, the
cooling laser will continue to drive the core transition so that the atom
will be in the core-excited state for a significant fraction of time.
Because the energy connected with the core transition is of the
order of $10^4$ K, each free electron produced by autoionization will rapidly
leave the plasma volume. Hence, the combined action of the
cooling laser, TBR and collisional deexcitation is expected to
remove electrons from the plasma and destroy the plasma, until recombination
stops when the electron density has become too small. In
order to suppress this electron loss one has to choose initial conditions
which lead to a tolerable TBR rate \cite{Man69}. 

\begin{figure}[tb]
\centerline{\psfig{figure=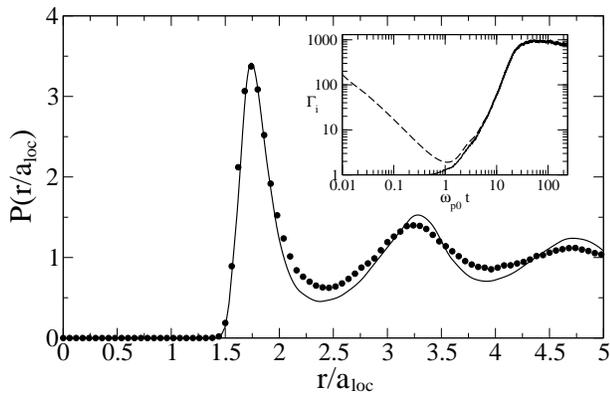,width=8cm}}
\caption{\label{fig2}Distribution of scaled inter-ionic distances after a time
of $t=52 \: \mu$s ($\omega_{p0}t=240$) compared to the calculated
pair-correlation function of an OCP at $\Gamma_i=700$ \cite{Ng74}. The inset shows the
ionic CCP calculated by different methods (see text for details).}
\end{figure}

If time and length are measured in units of the initial ion plasma frequency $\omega_{p0}$
and Wigner-Seitz radius $a_0$, respectively, the initial plasma state is
characterized by five parameters:
the number of ions $N_i$, the initial electron and ion CCPs,
the value of $\Gamma_c$ corresponding to the Doppler temperature
$T_c$, and
the ratio of the cooling rate to initial ionic plasma frequency $\omega_{p0}$. The exact
values of $T_c$ and $T_i(0)$ are not important for the plasma
expansion dynamics, since both are negligible compared to the electron
temperature. According to Eq.~(\ref{quneutr}), for fixed $\beta$ the time evolution of the scaled
density only depends on the product of $\Gamma_e$ and $N_{i}^{2/3}$
(if $T_i \ll T_e$ is neglected in the second equation of Eq.\ (\ref{quneutr})),
the expansion being slower if $\Gamma_e N_{i}^{2/3}$ becomes larger.
Hence, in order to slow down the plasma expansion sufficiently for spatial
correlations to develop, one may prefer to increase the ion number
since reduction of the  electron temperature becomes ultimately incompatible with the
objective of limiting TBR.
A further constraint on the initial conditions arises from the fact that
$\Gamma_i\approx 10^3$ in order to observe ordering effects. Typically, ion
temperatures of the order of 1 mK can be achieved through Doppler
cooling, meaning that the initial ion density must be at least
about $10^{8}$ cm$^{-3}$. 
Finally, the cooling rate $\beta$ must be of the order of $10^{-1}\omega_{p0}$
to sufficiently slow down the plasma expansion that correlations can
develop. Since $\beta$\ ($ \propto 1/m_i$) decreases faster than $\omega_{p0}$\ ($
\propto 1/\sqrt{m_i}$) with increasing ion mass, it is advantageous to consider
relatively light ions, for which sufficiently high cooling rates can be
obtained experimentally.
We therefore choose Be ions for our simulations, for which
laser-cooling has been experimentally demonstrated earlier in nonneutral
plasmas \cite{Bre88}.

\begin{figure}[b]
\centerline{\psfig{figure=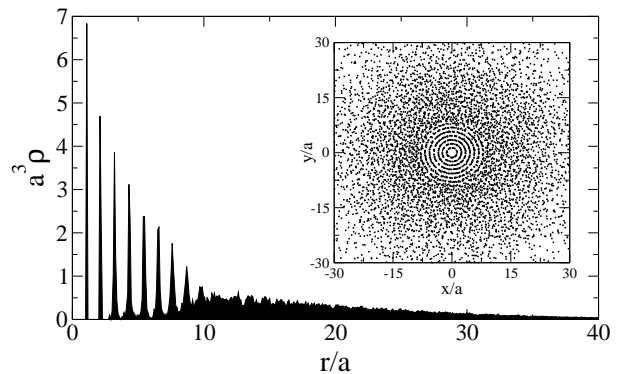,width=8cm}}
\caption{\label{fig3}Radial density after a time of $t = 24 \: \mu$s
($\omega_{p0} t=110$). The inset shows a two-dimensional cut through the plasma
cloud, clearly revealing the formation of concentric shells. (For better
contrast, cuts with $x=0$, $y=0$ and $z=0$, respectively, have been overlayed.)}
\end{figure}

The most striking result of our simulations  is a lattice-type crystallization of the ions
or even their arrangement in concentric shells if the plasma
expansion is slow enough. The emergence of such an order depends sensitively on the initial conditions.
 We have found a lattice-type crystallization in 
the expansion of a plasma of $20000$ Be-ions with an initial density of $1.1
\cdot 10^8$ cm$^{-3}$ and an electron temperature of $T_e=29$ K, cooled with a damping rate
of $\beta=0.15 \: \omega_{p0}$ to an ionic temperature of $T_c=2$ mK. The value of
$\Gamma_i=750$ after $52\mu$s, simply calculated from the ion temperature
and the average Wigner-Seitz radius, suggests strong ordering of the ionic
component. However, as pointed out in \cite{Kuz02} the CCP calculated in this way
may have no meaning as a measure of correlations, since the expanding
plasma does not reach a global equilibrium. A more reliable quantity, namely
the distribution of inter-ionic distances, is shown in Fig.~\ref{fig2}. In order to account for the
nonhomogeneity of the plasma we have scaled the inter-particle spacing by the
Wigner-Seitz radius $a_{loc}$ determined
by the local density between the corresponding particles. For comparison,
the pair-correlation function obtained from the HNC-equations of a homogeneous
one-component plasma \cite{Ng74} is also shown. (However,
the distribution function shown in Fig.~\ref{fig2} should not be understood as
a pair-correlation function in a strict sense, since in the present case the
plasma is neither isotropic nor homogeneous.) The remarkable agreement shows
that under present conditions the calculated CCP indeed indicates the
degree of order in the expanding plasma and that the system has reached a
state of local equilibrium far beyond the known crystallization limit of
$\Gamma_i\approx174$ \cite{Dub99}. In order to study the dynamics of the crystallization
process we have determined $\Gamma_i$ also from the numerically obtained
average correlation energy together with an analytical approximation for this
quantity \cite{Poh03}. A comparison of the CCPs calculated by both methods is
shown in the inset of Fig.~\ref{fig2}. Initially there are large deviations
between both calculations reflecting the nonequilibrium character of the early
plasma state. However, after some inverse plasma
frequencies, the ion system reaches a local equilibrium and the CCPs obtained
from the different methods become identical. At longer times both curves may
diverge again due to a freezing out of ordered structures when the density becomes too small.

\begin{figure}[tb]
\centerline{\psfig{figure=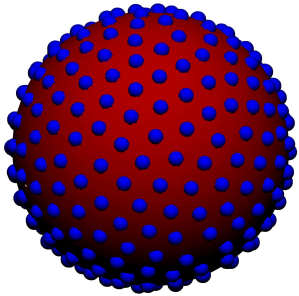,width=4cm}}
\caption{\label{fig4}(Color online) The 5$^{\text{th}}$ shell of Fig.~\ref{fig3}, demonstrating
significant intra-shell ordering.}
\end{figure}

If we further slow down the plasma expansion, by increasing the product $\Gamma_e N_{i}^{2/3}$, the system exhibits formation of concentric shells
rather than relaxation into lattice-type structures. This is demonstrated in
Fig.~\ref{fig3}, where the radial ion density is shown at $t=24\mu$s for a
plasma with $\Gamma_e=0.15$ and $N_i=50000$ while all other parameters
equal those used for Fig.~\ref{fig2}. Besides this radial ordering into concentric
shells, there are strong intra-shell correlations, also found in trapped
nonneutral plasmas \cite{Dub88}. In Fig.~\ref{fig4},
we show one of the shells formed in the simulation of
Fig.~\ref{fig3}. The development of a hexagonal-lattice like ordering is
evident, which is however considerably disturbed by the curvature of the shell.
A closer look on the emergence of the order shows that
in the early stages of the plasma evolution a cubic-lattice like structure is
formed. However, after some ten inverse plasma frequencies the ions rearrange
to form concentric shells starting from the plasma center, where the density
is highest, in contrast to trapped nonneutral plasmas where the shell
formation was observed to proceed from the periphery to the center of the cloud
\cite{Tot02}. If the expansion is faster, as in the first example
(Fig.~\ref{fig2}), ion-ion collisions are less frequent and after the initial
phase of local ordering the
density is already too low for the rearrangement into shells, such
that the lattice structure survives during the expansion.  

In summary, we have followed the long-time dynamics of
laser-cooled, expanding ultracold plasmas on the basis of a hybrid-MD
simulation, allowing for a full treatment of the strongly coupled ion
dynamics. The results show that cooling during the plasma expansion drastically modifies the
expansion dynamics leading to an exotic type of plasma where the electron component
 is weakly coupled while the ion component shows strong coupling
effects which manifest themselves in the development of lattice-like
structures (short-range order) or even 
the formation of concentric shells (long-range order) depending on the expansion dynamics. 
Interesting questions concerning this novel
system, like the behavior of ion collective modes or the influence of
different density profiles on the details of the structure formation process,
have to be addressed in future studies. 

Financial support from the DFG through grant RO1157/4 is gratefully
acknowledged.


\begin{thebibliography}{21}
\expandafter\ifx\csname natexlab\endcsname\relax\def\natexlab#1{#1}\fi
\expandafter\ifx\csname bibnamefont\endcsname\relax
  \def\bibnamefont#1{#1}\fi
\expandafter\ifx\csname bibfnamefont\endcsname\relax
  \def\bibfnamefont#1{#1}\fi
\expandafter\ifx\csname citenamefont\endcsname\relax
  \def\citenamefont#1{#1}\fi
\expandafter\ifx\csname url\endcsname\relax
  \def\url#1{\texttt{#1}}\fi
\expandafter\ifx\csname urlprefix\endcsname\relax\def\urlprefix{URL }\fi
\providecommand{\bibinfo}[2]{#2}
\providecommand{\eprint}[2][]{\url{#2}}

\bibitem[{\citenamefont{Rahman and Schiffer}(1986)\citenamefont{Rahman and
Schiffer}}]{Rah86}
\bibinfo{author}{\bibfnamefont{A.}~\bibnamefont{Rahman}} \bibnamefont{and}
  \bibinfo{author}{\bibfnamefont{J.P.}~\bibnamefont{Schiffer}},
  \bibinfo{journal}{Phys.\ Rev.\ Lett.} \textbf{\bibinfo{volume}{57}},
  \bibinfo{pages}{1133} (\bibinfo{year}{1986}).

\bibitem[{\citenamefont{Dubin and O'Neil}(1988)\citenamefont{Dubin and O'Neil}}]{Dub88}
\bibinfo{author}{\bibfnamefont{D.H.E.}~\bibnamefont{Dubin}} \bibnamefont{and}
  \bibinfo{author}{\bibfnamefont{T.M.}~\bibnamefont{O'Neil}},
  \bibinfo{journal}{Phys.\ Rev.\ Lett.} \textbf{\bibinfo{volume}{60}},
  \bibinfo{pages}{511} (\bibinfo{year}{1988}).

\bibitem[{\citenamefont{Gilbert, Bollinger and Wineland}(1988)
\citenamefont{Gilbert, Bollinger and Wineland}}]{Gil88}
  \bibinfo{author}{\bibfnamefont{S.L.}~\bibnamefont{Gilbert}},
  \bibinfo{author}{\bibfnamefont{J.J.}~\bibnamefont{Bollinger}},
 \bibnamefont{and}
  \bibinfo{author}{\bibfnamefont{D.J.}~\bibnamefont{Wineland}},
  \bibinfo{journal}{Phys.\ Rev.\ Lett.} \textbf{\bibinfo{volume}{60}},
  \bibinfo{pages}{2022} (\bibinfo{year}{1988}).

\bibitem[{\citenamefont{Dubin and O'Neil}(1999)\citenamefont{Dubin and O'Neil}}]{Dub99}
\bibinfo{author}{\bibfnamefont{D.H.E.}~\bibnamefont{Dubin}} \bibnamefont{and}
  \bibinfo{author}{\bibfnamefont{T.M.}~\bibnamefont{O'Neil}},
  \bibinfo{journal}{ Rev.\ Mod.\ Phys.} \textbf{\bibinfo{volume}{71}},
  \bibinfo{pages}{87} (\bibinfo{year}{1999}).

\bibitem[{\citenamefont{Totsuji, Kishimoto, Totsuji and Tsurata}(2002)}]{Tot02}
\bibinfo{author}{\bibfnamefont{H.}~\bibnamefont{Totsuji}},
\bibinfo{author}{\bibfnamefont{T.}~\bibnamefont{Kishimoto}},
\bibinfo{author}{\bibfnamefont{C.}~\bibnamefont{Totsuji}},
\bibnamefont{and}
  \bibinfo{author}{\bibfnamefont{K.}~\bibnamefont{Tsuruta}},
  \bibinfo{journal}{Phys.\ Rev.\ Lett.} \textbf{\bibinfo{volume}{88}},
  \bibinfo{pages}{125002} (\bibinfo{year}{2002}).

\bibitem[{\citenamefont{Killian et~al.}(1999)\citenamefont{Killian, Kulin,
  Bergeson, Orozco, Orzel, and Rolston}}]{Kil99}
\bibinfo{author}{\bibfnamefont{T.C.}~\bibnamefont{Killian}},
  \bibinfo{author}{\bibfnamefont{S.}~\bibnamefont{Kulin}},
  \bibinfo{author}{\bibfnamefont{S.D.}~\bibnamefont{Bergeson}},
  \bibinfo{author}{\bibfnamefont{L.A.}~\bibnamefont{Orozco}},
  \bibinfo{author}{\bibfnamefont{C.}~\bibnamefont{Orzel}}, \bibnamefont{and}
  \bibinfo{author}{\bibfnamefont{S.L.}~\bibnamefont{Rolston}},
  \bibinfo{journal}{Phys.\ Rev.\ Lett.} \textbf{\bibinfo{volume}{83}},
  \bibinfo{pages}{4776} (\bibinfo{year}{1999}).

\bibitem[{\citenamefont{Kulin et~al.}(2000)\citenamefont{Kulin, Killian,
  Bergeson, and Rolston}}]{Kul00}
\bibinfo{author}{\bibfnamefont{S.}~\bibnamefont{Kulin}},
  \bibinfo{author}{\bibfnamefont{T.C.}~\bibnamefont{Killian}},
  \bibinfo{author}{\bibfnamefont{S.D.}~\bibnamefont{Bergeson}},
  \bibnamefont{and}
  \bibinfo{author}{\bibfnamefont{S.L.}~\bibnamefont{Rolston}},
  \bibinfo{journal}{Phys.\ Rev.\ Lett.} \textbf{\bibinfo{volume}{85}},
  \bibinfo{pages}{318} (\bibinfo{year}{2000}).

\bibitem[{\citenamefont{Killian et~al.}(2001)\citenamefont{Killian, Lim, Kulin,
  Dumke, Bergeson, and Rolston}}]{Kil01}
\bibinfo{author}{\bibfnamefont{T.C.}~\bibnamefont{Killian}},
  \bibinfo{author}{\bibfnamefont{M.J.}~\bibnamefont{Lim}},
  \bibinfo{author}{\bibfnamefont{S.}~\bibnamefont{Kulin}},
  \bibinfo{author}{\bibfnamefont{R.}~\bibnamefont{Dumke}},
  \bibinfo{author}{\bibfnamefont{S.D.}~\bibnamefont{Bergeson}},
  \bibnamefont{and}
  \bibinfo{author}{\bibfnamefont{S.L.}~\bibnamefont{Rolston}},
  \bibinfo{journal}{Phys.\ Rev.\ Lett.} \textbf{\bibinfo{volume}{86}},
  \bibinfo{pages}{3759} (\bibinfo{year}{2001}).

\bibitem[{\citenamefont{Kuzmin and O'Neil}(2002)}]{Kuz02}
\bibinfo{author}{\bibfnamefont{S.G.}~\bibnamefont{Kuzmin}} \bibnamefont{and}
  \bibinfo{author}{\bibfnamefont{T.M.}~\bibnamefont{O'Neil}},
  \bibinfo{journal}{Phys.\ Rev.\ Lett.} \textbf{\bibinfo{volume}{88}},
  \bibinfo{pages}{065003} (\bibinfo{year}{2002}).

\bibitem[{\citenamefont{Murillo}(2001)}]{Mur01}
\bibinfo{author}{\bibfnamefont{M.S.}~\bibnamefont{Murillo}},
  \bibinfo{journal}{Phys.\ Rev.\ Lett.} \textbf{\bibinfo{volume}{87}},
  \bibinfo{pages}{115003} (\bibinfo{year}{2001}).

\bibitem[{\citenamefont{Robicheaux and Hanson}(2002)}]{Rob02}
\bibinfo{author}{\bibfnamefont{F.}~\bibnamefont{Robicheaux}} \bibnamefont{and}
  \bibinfo{author}{\bibfnamefont{J.D.}~\bibnamefont{Hanson}},
  \bibinfo{journal}{Phys.\ Rev.\ Lett.} \textbf{\bibinfo{volume}{88}},
  \bibinfo{pages}{055002} (\bibinfo{year}{2002}).

\bibitem[{\citenamefont{Robicheaux et~al.}(2003)\citenamefont{Robicheaux and Hanson}}]{Rob03}
\bibinfo{author}{\bibfnamefont{F.}~\bibnamefont{Robicheaux}}
\bibnamefont{and}	
  \bibinfo{author}{\bibfnamefont{J.D.}~\bibnamefont{Hanson}},
  \bibinfo{journal}{Phys.\ Plasmas} \textbf{\bibinfo{volume}{10}},
  \bibinfo{pages}{2217} (\bibinfo{year}{2003}).

\bibitem[{\citenamefont{Kuzmin et~al.}(2002)\citenamefont{Kuzmin and O'Neil}}]{Kuz02b}
\bibinfo{author}{\bibfnamefont{S.G.}~\bibnamefont{Kuzmin}}
\bibnamefont{and}	
  \bibinfo{author}{\bibfnamefont{T.M.}~\bibnamefont{O'Neil}},
  \bibinfo{journal}{Phys.\ Plasmas} \textbf{\bibinfo{volume}{9}},
  \bibinfo{pages}{3743} (\bibinfo{year}{2002}).

\bibitem[{\citenamefont{Killian, Ashoka, Gupta, Laha, Nagel, Simien, Kulin, Rolston and Bergeson}(2003)}]{Kil03}
\bibinfo{author}{\bibfnamefont{T.C.}~\bibnamefont{Killian}},
\bibinfo{author}{\bibfnamefont{V.S.}~\bibnamefont{Ashoka}},
\bibinfo{author}{\bibfnamefont{P.}~\bibnamefont{Gupta}},
\bibinfo{author}{\bibfnamefont{S.}~\bibnamefont{Laha}},
\bibinfo{author}{\bibfnamefont{S.B.}~\bibnamefont{Nagel}},
\bibinfo{author}{\bibfnamefont{C.E.}~\bibnamefont{Simien}},
\bibinfo{author}{\bibfnamefont{S.}~\bibnamefont{Kulin}},
\bibinfo{author}{\bibfnamefont{S.L.}~\bibnamefont{Rolston}},
\bibnamefont{and}
\bibinfo{author}{\bibfnamefont{S.D.}~\bibnamefont{Bergeson}},
  \bibinfo{journal}{J.\ Phys.~A} \textbf{\bibinfo{volume}{36}},
  \bibinfo{pages}{6077} (\bibinfo{year}{2003}).

\bibitem[{\citenamefont{Dorozhkina and Semenov}(1998)}]{Dor98}
\bibinfo{author}{\bibfnamefont{D.S.}~\bibnamefont{Dorozhkina}} \bibnamefont{and}
  \bibinfo{author}{\bibfnamefont{V.E.}~\bibnamefont{Semenov}},
  \bibinfo{journal}{Phys.\ Rev.\ Lett.} \textbf{\bibinfo{volume}{81}},
  \bibinfo{pages}{2691} (\bibinfo{year}{1998}).

\bibitem[{\citenamefont{Metcalf and van der Straten}(1999)}]{Met99}
\bibinfo{author}{\bibfnamefont{H.J.}~\bibnamefont{Metcalf}}
\bibnamefont{and}
\bibinfo{author}{\bibfnamefont{P.}~\bibnamefont{van der Straten}},
  \emph{\bibinfo{title}{Laser Cooling and Trapping}}
(\bibinfo{publisher}{Springer, New York}, \bibinfo{year}{1999}).

\bibitem[{\citenamefont{Pohl et~al.}()\citenamefont{Pohl, Pattard, and
  Rost}}]{Poh03}
\bibinfo{author}{\bibfnamefont{T.}~\bibnamefont{Pohl}},
  \bibinfo{author}{\bibfnamefont{T.}~\bibnamefont{Pattard}}, 
\bibnamefont{and}
  \bibinfo{author}{\bibfnamefont{J.M.}~\bibnamefont{Rost}},
  \bibinfo{journal}{Phys.\ Rev.\ A} \textbf{\bibinfo{volume}{68}},
  \bibinfo{pages}{010703(R)} (\bibinfo{year}{2003}).

\bibitem[{\citenamefont{Barnes and Hut}(1986)}]{Bar86}
\bibinfo{author}{\bibfnamefont{J.}~\bibnamefont{Barnes}} \bibnamefont{and}
  \bibinfo{author}{\bibfnamefont{P.}~\bibnamefont{Hut}},
  \bibinfo{journal}{Nature} \textbf{\bibinfo{volume}{324}},
  \bibinfo{pages}{446} (\bibinfo{year}{1986}).

\bibitem[{\citenamefont{Mazevet et~al.}(2002)\citenamefont{Mazevet, Collins,
  and Kress}}]{Maz02}
\bibinfo{author}{\bibfnamefont{S.}~\bibnamefont{Mazevet}},
  \bibinfo{author}{\bibfnamefont{L.A.}~\bibnamefont{Collins}},
\bibnamefont{and}
  \bibinfo{author}{\bibfnamefont{J.D.}~\bibnamefont{Kress}},
  \bibinfo{journal}{Phys.\ Rev.\ Lett.} \textbf{\bibinfo{volume}{88}},
  \bibinfo{pages}{055001} (\bibinfo{year}{2002}).

\bibitem[{\citenamefont{Birdsall}(1991)}]{Vah95}
\bibinfo{author}{\bibfnamefont{C.K.}~\bibnamefont{Birdsall}},
  \bibinfo{journal}{IEEE\ Trans.\ Plasma\ Sci.} \textbf{\bibinfo{volume}{19}},
  \bibinfo{pages}{65} (\bibinfo{year}{1991}).

\bibitem[{\citenamefont{Gallagher}(1994)}]{Gal94}
\bibinfo{author}{\bibfnamefont{T.F.}~\bibnamefont{Gallagher}},
  \emph{\bibinfo{title}{Rydberg Atoms}} (\bibinfo{publisher}{Cambridge
  University Press}, \bibinfo{year}{1994}).

\bibitem[{\citenamefont{Mansbach and Keck}(1969)}]{Man69}
\bibinfo{author}{\bibfnamefont{P.}~\bibnamefont{Mansbach}} \bibnamefont{and}
  \bibinfo{author}{\bibfnamefont{J.}~\bibnamefont{Keck}},
  \bibinfo{journal}{Phys.\ Rev.} \textbf{\bibinfo{volume}{181}},
  \bibinfo{pages}{275} (\bibinfo{year}{1969}).

\bibitem[{\citenamefont{Brewer, Prestage, Bollinger, Itano, Larson and Wineland}(1988)}]{Bre88}
\bibinfo{author}{\bibfnamefont{L.R.}~\bibnamefont{Brewer}},
\bibinfo{author}{\bibfnamefont{J.D.}~\bibnamefont{Prestage}},
\bibinfo{author}{\bibfnamefont{J.J.}~\bibnamefont{Bollinger}},
\bibinfo{author}{\bibfnamefont{W.M.}~\bibnamefont{Itano}},
\bibinfo{author}{\bibfnamefont{D.J.}~\bibnamefont{Larson}},
\bibnamefont{and}
\bibinfo{author}{\bibfnamefont{D.J.}~\bibnamefont{Wineland}},
  \bibinfo{journal}{Phys.\ Rev.\ A} \textbf{\bibinfo{volume}{38}},
  \bibinfo{pages}{859} (\bibinfo{year}{1988}).

\bibitem[{\citenamefont{Ng}(1974)}]{Ng74}
\bibinfo{author}{\bibfnamefont{K.C.}~\bibnamefont{Ng}},
  \bibinfo{journal}{J.\ Chem.\ Phys.} \textbf{\bibinfo{volume}{61}},
  \bibinfo{pages}{2680} (\bibinfo{year}{1974}).

\end{thebibliography}
\end{document}